\documentclass[final,5p,times,twocolumn]{elsarticle}

\usepackage{amssymb}
\usepackage{longtable}
\usepackage{graphicx}
\usepackage{graphics}
\usepackage{bbm}
\usepackage{amsmath}

\usepackage[normalem]{ulem}
\usepackage{color}

\begin{document}

\begin{frontmatter}



\title{Quantum Dissipation in a Neutrino System Propagating in Vacuum and in Matter}

\author[label1]{Marcelo M. Guzzo}
\author[label1]{Pedro C. de Holanda}
\author[label1,label2]{Roberto L. N. Oliveira}
\address[label1]{Instituto de F\'\i sica Gleb Wataghin 
Universidade Estadual de Campinas, UNICAMP 
13083-970, Campinas, S\~ao Paulo, Brasil}
\address[label2]{Northwestern University, Department of Physics \& Astronomy, 2145 Sheridan Road, Evanston, IL 60208, USA}

\begin{abstract}

  Considering the neutrino state like an open quantum system, we analyze its propagation in vacuum or in matter. After defining  what can be called decoherence and relaxation effects, we show that in general the probabilities in vacuum and in constant matter can be written in a similar way, which is not an obvious result for such system. From this result, we analyze the situation where neutrino evolution satisfies the adiabatic limit and use this formalism to study solar neutrinos. We show that the decoherence effect may not be bounded  by the solar neutrino data and review some results in the literature, in particular the current results where solar neutrinos were used to put bounds on decoherence effects through a model-dependent approach. We conclude explaining how and why these models are not general and we reinterpret these constraints.

\end{abstract}


\end{frontmatter}


\section{Introduction}

We present a study on dissipative effects on neutrino evolution, such as the decoherence and relaxation effects, and their consequences in neutrino oscillations. These effects are obtained when we consider neutrinos as an open quantum system~\cite{ben,ben1,workneut}. In this approach, neutrinos are considered as a subsystem that is free to interact with the environment that presents a reservoir behavior.~\footnote{Some possible sources of violations of quantum mechanics fundamentals include the spontaneous evolution of pure states into mixed decoherent states~\cite{1} induced by interactions with the space-time at Planck scale~\cite{2} which unavoidly appear in any formulation of a quantum gravity theory. Such sources of decoherence was first analyzed in  Ref.~\cite{3} which considered oscillating systems propagating over large distances and the corresponding damping effects in the usual interferometric pattern characterizing the oscillation phenomenon.}

The decoherence effect is the most usual dissipative effect. In the neutrino oscillation phenomenon, the decoherence effect acts only on the quantum interference, dynamically eliminating the oscillating terms in oscillation probabilities. This feature has 
been investigated in a number of previous studies~\cite{dea,lis,fun1,gab,fo,liss,mmy,workneut3}.

The relaxation effect acts in a different way and it does not affect the oscillating terms. It changes only the pure mixing terms in the probabilities, leading all averaged conversion probabilities to $1/n$, where $n$ is the number of neutrino families. Then, the relaxation effect can change the probability behavior even when the oscillation terms are not important,  like the solar neutrino case \cite{workneut}.
The relaxation effect can be confused with the decoherence effect and this can occur in those particular cases where quantum coherence is averaged out in neutrino oscillations. In Ref.~\cite{fo}, the authors analyzed quantum decoherence effect with solar and KamLAND neutrinos. However, for solar neutrinos the decoherence effect could be investigated only using a model-dependent approach, because in general, the quantum coherence is averaged out for solar neutrinos and just relaxation effects can be investigated. 

There are some experimental bounds on dissipative effects and we will compare some concrete bounds obtained from some experimental data analyses found in the literature. All these limits were obtained for neutrino propagation in vacuum and in two neutrino approximation. For example, in Ref.~\cite{workneut3}, the analysis was made considering MINOS experiment. There, the  decoherence parameter has a superior limit given by $\gamma<9.11 \times 10^{-23}$ GeV at $95\%$ C.L. and this result agrees with the upper limit found in Ref.~\cite{lis} where $\gamma<4.10 \times 10^{-23}$ GeV at $95 \%$ C.L., which was obtained for atmospheric neutrino case. 

A very interesting upper limit was introduced by Ref. \cite{fo}  obtained in a model-dependent approach that constrain decoherence effect using solar neutrinos. It was obtained that decoherence parameter is limited to $\gamma<0.64\times 10^{-24}$ GeV at $95\%$ C.L. As it is known, the matter effect is important in this case, and we will address this issue later on this article. In~\cite{balieiro} an analysis using only reactor neutrinos found different bounds on the decoherence effect,  $\gamma<{6.8\times 10^{-22}}$ eV at $95\%$ at C.L. All bounds presented above can be found in Table \ref{table1}.\footnote{Following the arguments of the present article, decoherence effect can be described by one parameter and relaxation effect by another parameter. However, in the case of three neutrino oscillation there are three different decoherence parameters and two different relaxation parameters. As we can see in Ref.~\cite{workneut2}, the decoherence parameters describe the quantum effect between specific families and then, the decoherence bound for accelerator or atmospheric neutrinos can be different from the one for reactor neutrinos.}

In general, bounds on dissipative parameters come from $e^{-\gamma x}\lesssim 1$ since this is the kind of damping terms which  appear in the oscillation probabilities. This can be checked to work reasonably well for all the limits presented above, with terrestrial experiments with a typical baseline $x=10^{20}\sim10^{22}$ GeV$^{-1}$ (20$\sim$ 2000 km).

However for the numbers presented in \cite{fo}, using the bound found for $\gamma< 0.64\times 10^{-24}$ GeV, the exponential term tends strongly to $1$.
As it will be clear in this work, the model-dependent approach used in Ref. \cite{fo} also constrains the relaxation effect with $\gamma_{relax.}<10^{-25}$ GeV at $95\%$ C.L. For solar-neutrinos $x=10^{26}$ GeV$^{-1}$, and the exponential term in this case makes the survival probability for solar neutrinos to have a unique constant value equal to $1/2$. This result should spoil the usual solution for solar neutrinos.  In our model, the constraint for $\gamma$ is expected to be two order of magnitude smaller \cite{escrever}.

In the particular case investigated in Ref.~\cite{fo}, where this limit was obtained in a model-dependent approach, the exponential argument depends on other oscillation parameters, including necessarily the neutrino energy, and this makes the bound on $\gamma$ just suitable in that situation.

In the model-independent approach that we will introduce in this work, the damping term will not depend on any oscillation parameters and the addition of any energy dependence on $\gamma$ will be an ansatz, as those found in Refs. \cite{lis,fo,workneut3}. Besides, in our model the damping term for solar neutrino does not describe the decoherence effect, but only the relaxation effect. In fact, following the definitions that we will present in this work, the bound found in Ref. \cite{fo} can be called of decoherence just because it is proportional to the relaxation effect, which is, in fact, the only dissipative effect that remains after averaging out the solar-neutrino oscillations. Furthermore, our model respects the usual bound condition $(e^{-x \gamma} \lesssim 1)$ for the damping terms in the neutrino probabilities.

Our analysis will consider these two non-standard effects. We analyze the propagation in vacuum and in matter. We show that with a careful application of  the open quantum system theory it is possible to write the probabilities in vacuum and in constant matter in a similar way, which is not an obvious result in this context. From this result, we analyze the situation where neutrino evolution satisfies the adiabatic limit, and analyze solar neutrinos in two neutrino approximation to show that the decoherence effect cannot be bounded in general using this neutrino source~\cite {farzan}. We discuss the current results \cite{fo, bur} where solar neutrinos were used to put limits on decoherence effect through a model-dependent approach. We argue how and why these models are not general and we reinterpret these constraints.

We conclude this work arguing that the decoherence limit in the channel $\nu_{e}\rightarrow \nu{_\mu}$ can be different from the limit obtained in Ref.~\cite{fo}. A limit for decoherence parameters can be obtained using a model-independent approach studying neutrinos from sources other than the sun.

\begin{center}
\begin{table}[!t]
\caption{\label{table1}Upper limits on decoherence parameters at $95\%$ C. L. obtained from accelerator, atmospheric, reactor and solar experiments, respectively. These  bounds assume that the decoherence parameters are energy independent. }
\vspace{0.5 cm}
\hspace{0.5cm}\begin{tabular}{||c||c||c||}
\hline
\hline
$P(\nu_\alpha \nu_\alpha)$ & $\gamma$ in GeV & baseline/E \\
\hline
$P(\nu_\mu \nu_\mu)$ & $9.11 \times 10^{-23}$~\cite{workneut3}& $\sim 730$ km $/3$ GeV \\
\hline
$P(\nu_\mu \nu_\mu)$     & $4.10 \times 10^{-23}$~\cite{lis}     & $\lesssim 10^{4}$ km $/ 10^{3}$ GeV\\
\hline
$P(\bar{\nu}_e\bar{\nu}_e)$ &  $6.8\times 10^{-22}$~\cite{balieiro} & $\sim 200$ km $/5$ MeV\\
\hline
$P(\nu_e\nu_e)$ &  $0.64\times 10^{-24}$~\cite{fo} & $\sim 10^{8}$  km $/ 2$ MeV \\
\hline
\hline
\end{tabular}
\end{table}
\end{center}
\vspace{-0.5 cm}

\section{Neutrinos as an Open Quantum System}

In open quantum system approach, a global state formed by a subsystem of interest and an environment must be defined. As the environment in this approach is a quantum reservoir, it interacts with the subsystem of interest as a whole. 


The subsystem of interest can be represented by $S$ states which are associated with the Hilbert space $\mathbbm{H}_{S}$,  while the quantum reservoir can be represented by $R$ states which are associated with the Hilbert space $\mathbbm{H}_{R}$. Basically, those are the fundamental definition about these two different quantum states.  The subsystem of interest may be composed by more than one Hilbert space associated with each element that can be added in the usual quantum description of a system. For instance, when the matter potential is added to mass Hamiltonian in neutrino oscillation in vacuum. 

The product tensor from these spaces form the total Hilbert space or the global states space, $\mathbbm{H}_{G}=\mathbbm{H}_{S}\otimes\mathbbm{H}_{R}$. This means that we can write a global state as~\cite{len, pet}
\begin{equation}
\rho_{G} = \rho_{S}\otimes \omega_{R}\,,
\label{i} 
\end{equation}
where $\rho_{S}$ is the subsystem of interest state, and $\omega_{R}$ is the reservoir state. The system evolution is obtained using the following transformation:  
\begin{equation}
\rho_{G}(t) = U(\rho_{S}\otimes \omega_{R})U^{\dag},
\label{ii} 
\end{equation}
such that $U=Exp[-i H_{tot} t]$ is the unitary operator and the time evolution is governed by the total Hamiltonian that can be defined as $H_{tot}=H_{S}+H_{R}+H_{int}$, where $H_{S}$ is the subsystem of interest Hamiltonian, $H_{R}$ is the reservoir Hamiltonian and $H_{int}$ is the interaction Hamiltonian between reservoir and subsystem of interest.

The subsystem of interest changes its characteristic in time due to the internal dynamic and  the interaction with the reservoir~\cite{len, pet}. On the other hand, as that reservoir state does not change in time, its dynamics is not important. Then, the dynamic of the subsystem of interest  is obtained taking the trace over the reservoir states in Eq. (\ref{ii})~\cite{lin,dav,dum,kra}, i. e.,
\begin{equation}
\rho_{S}(0)\rightarrow \rho_{s}(t)=\Lambda \rho_{s}(0)=Tr_{R} U(\rho_{S}\otimes \omega_{R})U^{\dag},
\label{iii} 
\end{equation}
where $\Lambda$ is a dynamic map. Eq. (\ref{iii}) is known as the reduced dynamic of $S$.  Solving the partial trace in Eq. (\ref{iii}), we can rewrite this relation as   
\begin{equation}
\Lambda \rho_{S}(0)=\sum_{\alpha} W_{\alpha}\rho_{S} W^{\dag	}_{\alpha},
\label{iv} 
\end{equation}
where $W_{\alpha}\in \mathbbm{H}_{S}$ and $\sum_{\alpha} W_{\alpha}W^{\dag}_{\alpha}=\mathbbm 1$~\cite{kra}. In order to evolve the state, this map must satisfy the complete positivity constraint. Besides, we need a family of linear maps which must satisfy the semigroup properties \cite{lin,dav,kra}. From this, we can obtain a dynamical generator, which can be written as
\begin{equation}
\frac{d\rho_{\nu}(t)}{dt}=-i[H_{S},\rho_{\nu}(t)]+D[\rho_{\nu}(t)]\,.
\label{v}
\end{equation}  

This equation has been studied in literature and more information about it and its properties can be found in Refs.~\cite{len,pet,lin,dav, dum,kra,joo,uri}. This equation is called Lindblad Master Equation and it is composed by an usual Hamiltonian term and a non-Hamiltonian one which gives origin to dissipative effects. The dissipator in Eq. (\ref{v}) can be defined as
\begin{equation}
D[\rho_{\nu}] =\frac{1}{2}\sum_{k=1}^{N^{2}-1}\bigg(\Big[V_{k},\rho_{\nu} V_{k}^{\dag}\Big]+\Big[V_{k}\rho_{\nu},V_{k}^{\dag}\Big]\bigg) \,,
\label{vi}
\end{equation}
where $V_{k}$ are dissipative operators which act only on the $N$-dimensional $\mathbbm{H}_{S}$ space. The trace preservation of $\rho_{\nu}$ occurs only if $\sum_{k} V^{\dag}_{k} V_{k} =1$ is satisfied. The $V_{k}$ operators arise from the interaction of the subsystem of interest with the environment. The propagation through equation (\ref{v}) leads an initial density matrix state into a new density matrix state~\cite{ben}. The evolution is complete positive, transforming pure states into mixed states due to dissipation effects \cite{len,lin,dav,dum,kra}. The Von Neumann entropy of the subsystem of interest, $S=-Tr[\rho_{\nu}ln \rho_{\nu}]$, must be increasing in time and this is guaranteed if we impose $V^{\dag}_{k}=V_{k}$ \cite{nar}.

Let us start considering only two neutrino families and the relation between the mass and flavor bases in vacuum is given by \cite{moh,kim}
\begin{equation}
\rho_{m}=U^{\dag}\rho_{f}U\,,
\label{vii}
\end{equation}
where $\rho_{m}$ is written in mass basis, $\rho_{f}$ is written in flavor basis and $U$ is the usual $2\times 2$ unitary mixing matrix.  

The transformation in Eq. (\ref{vii}) can be used to write the Eq. (\ref{v}) in the flavor basis or any other basis. Since  any unitary transformation over $V_{k}$, i. e., $A V_{k}A^{\dag}$ with $A A^{\dag}=1$, leads to a new matrix of the form:
\begin{equation}
V'_{k} = A V_{k}A^{\dag}= A \left(\begin{array}{ c c } 
V_{11} & V_{12}\\
V^{*}_{12} & V_{22}  \end{array} \right) A^{\dag} =
\left(\begin{array}{ c c } 
V'_{11} & V'_{12}\\
V'^{*}_{12} & V'_{22} \end{array} \right)\,,
\label{viiextra}
\end{equation}
where the new dissipator can be reparametrized such that it has the same form of the old dissipation operator. 


Expanding  Eqs.~(\ref{v}) and (\ref{vi}) in $SU(2)$ basis matrices we can write Eq. (\ref{v}) as:
\begin{equation}
\frac{d}{dx}\rho_{\mu}(x)\sigma_{\mu}=2\epsilon_{ijk}H_{i}\rho_{j}(x)\sigma_{\mu}\delta_{\mu k}+D_{\mu\nu}\rho_{\nu}(x)\sigma_{\mu}\,,
\label{viii}
\end{equation}
with $D_{\mu 0} =D_{0\nu}=0$ to keep the probability conservation. The matrix $D_{m n}$ can be parametrized  as
\begin{equation}
D_{mn} = -
\left(\begin{array}{ c c c} 
\gamma_{1} & \alpha & \beta\\
\alpha & \gamma_{2} & \delta\\
\beta & \delta & \gamma_{3} \end{array} \right)\,,
\label{ix}
\end{equation}
where the complete positivity constrains each parameter in the following form
\begin{eqnarray*}
2R  & \equiv & \gamma_{1}+\gamma_{2}-\gamma_{3} \geq 0;
\quad\mbox{ }\quad RS-\alpha^{2} \geq 0;\nonumber\\
2S & \equiv & \gamma_{1}+\gamma_{3}-\gamma_{2} \geq 0;
\quad\mbox{ }\quad  RT-\beta^{2}\geq 0;\nonumber\\
2T &\equiv & \gamma_{2}+\gamma_{3}-\gamma_{1} \geq 0;
\quad\mbox{ }\quad  ST-\delta^{2} \geq 0\,;
\label{iv.xx}   
\end{eqnarray*}
\begin{equation}
RST \geq 2\alpha\beta\delta+T\delta^{2}+S\beta^{2}+R\alpha^{2}\,.
\label{x}
\end{equation}
When we take out the reservoir Hamiltonian, $H_{R}$, and the interaction Hamiltonian, $H_{int}$,  the quantum evolution return to usual way and then the Eq. (\ref{v}), which is just the known Liouville quantum equation.

\subsection{The Subsystem of Interest}

Our subsystem of interest will be the neutrinos. As it is well known, many experiments give evidence that neutrinos have mass and mixing, as defined in Eq. (\ref{vii}), such that flavors oscillation can occur \cite{moh,kim}.

Neutrinos propagate in vacuum or in matter. In both situations it is possible to evolve neutrinos as an open quantum system, through direct application of the Eqs.~(\ref{v}) and (\ref{vi}). However, it is important to take into account in which circumstances these equations were developed and how the subsystem of interest was defined. Then, the definition of neutrinos like a subsystem of interest can change in each case. 

We can use a previous knowledge of the Hamiltonian in standard quantum mechanics to define this general subsystem of interest S. As we have seen, the total Hamiltonian in open quantum system approach can be defined as $H_{tot}=H_{S}+H_{R}+H_{int}$. In this case, $H_{S}$ is the usual Hamiltonian in closed approach. Then, the more general subsystem of interest is the physical object described by basis in which $H_{S}$ is diagonal.



\subsection{Quantum Dissipator and the Effects in $S$}

It is possible to study how each entry in the matrix in Eq. (\ref{ix}) changes the neutrino probabilities~\cite{workneut}. For simplicity  we will work with only two models for quantum dissipator. One with only one new parameter that will describes decoherence effect and another with two new different parameters that will describe decoherence and relaxation effects.

The most usual dissipator is obtained imposing energy conservation on the subsystem of interest $S$. This constraint satisfies the following commutation relation:  $[H_{S}, V_{k}]=0$. This dissipator adds only decoherence to the system of interest $S$ and it is given by 
\begin{equation}
D_{mn}=-diag\{\gamma_{1},\gamma_{1},0\} 
\label{xi}
\end{equation}
where, in this case $\gamma_{1}=\gamma_{2}$ and all other parameters vanish. This statement defines uniquely a particular interaction between the subsystem of interest $S$ and the reservoir. 

Therefore, the energy conservation constraint in subsystem of interest $S$ is obtained only if the commutation relation $[H_{S}, V_{k}]=0$ is satisfied and the consequence is a quantum dissipator with only one parameter, $\gamma_{1}$, that describes decoherence effect. In other words, the dynamic evolution is purely decoherent when this specific constraint is applied and no other dissipative effect is present.

To include the relaxation effect we need to violate the above constraint. As the subsystem of interest is free to interact with the reservoir the energy flux can fluctuate and the energy conservation condition imposed over the subsystem of interest can be not satisfied. In this case, the matrix in Eq. (\ref{ix}) can assume its complete form. However, as the matrix in Eq. (\ref{ix}) needs to be positive, all off-diagonal parameters must be smaller than the diagonal parameters. Then, only the diagonal parameter  necessarily must be present in case of new physics.  For simplicity, we will disregard all off-diagonal elements.

By assuming $[H_{S},V_{k}]\neq0$, a non null $D_{33}$ parameter can be included in the dissipator in Eq. (\ref{xi}) and then a new quantum dissipator can be written as

\begin{equation}
D_{mn}=-diag\{\gamma_{1},\gamma_{1},\gamma_{3}\}\,,
\label{xii}
\end{equation}
where $\gamma_{1}$ continues describing the decoherence effect and $\gamma_{3}$ describes the relaxation effect. 




\subsection{Dissipation in other Specific Subsystem of Interest S'}
\label{s}

The quantum dissipator written in Eq.~(\ref{vi}) can be defined in many different ways for neutrinos propagating in vacuum or in constant matter density, but it can have the same form in both cases. It is easy to prove this statement since we can always write $H_{S}$ in Eq. (\ref{ii}) as being diagonal in vacuum or in matter propagation. However, the parameter values in operator $V_{k}$ are different in each case. 

In the presence of matter the transformation between the effective mass basis and flavor basis can be written changing $\rho_{m}\rightarrow \tilde{\rho}_{m}$ and $U\rightarrow \tilde{U}$, where $\tilde{U}$ is composed by effective mixing angles \cite{kim, moh}. This transformation may not bring anything new to the quantum evolution equation in (\ref{v}) and it can be again parametrized as we made in Eq. (\ref{viii}) with a $\tilde{D}_{mn}$ that has the same form of the $D_{mn}$ that was given by Eq. (\ref{ix}).

In the usual situation in matter propagation we can define $H_{S}=H_{osc}+H_{mat}$ and then, the interaction constraints between a specific subsystem of interest $S'$ and the reservoir can be imposed in different ways.  Thus, it is possible to define a specific subsystem of interest $S'$ that can have a commutation relation with a particular $V_{k}$. While,  the $H_{S}$ defines the more general subsystem of interest S, the $ H_{osc}$ or $H_{mat}$ could be used to define other specific subsystems of interest S'. 

If we assume, for instance, that $[H_{osc},V_{k}]=0$, the energy conservation is kept when the propagation is in vacuum and only decoherence can act during the propagation. However, to the same case, when the propagation is in the matter $H_{S}\neq H_{osc}$ and therefore this constraint no longer preserve the energy conservation in the subsystem of interest $S$ and we have the situation where $[H_{S},V_{k}]\neq0$. Thus, the relaxation and decoherence effects may act during the propagation.

 
Therefore, when one defines $H_{S}$ and its relation with the $V_{k}$ operators, all the  dissipative effects are determined. So, a consequence of the definition of the subsystem of interest $S$ from $H_{S}$ can be summarized as follow: if the subsystem of interest $S$ has its energy conserved then $[H_{S},V_{k}]=0$ and the dissipator has the form of Eq. (\ref{xi}). In this case we are dealing with decoherence effects. If it is not, then $[H_{S},V_{k}]\neq0$ and the dissipator can be written in its more general form, Eq. (\ref{xii}). Thus, there are both decoherence and relaxation effects taking place during neutrino evolution.

The difference between decoherence and relaxation effect was discussed in this section. Now, we will apply this formalism in neutrino oscillation in vacuum and in constant matter case in order to eliminate any confusion between these two dissipative effects.

\section{Propagation  in Vacuum and in Constant Matter Density}

With the Lindblad Master Equation we can study many dissipative effects in neutrino oscillations. Decoherence is the most usual dissipative effect~\cite{fo,dea,gab,fun1,lis,liss,mmy,dan}, but it is not the only one, as we have seen in previous section. In particular, we are going to study how decoherence and relaxation effects act on the state during its propagation and how these dissipative effects change the oscillation probabilities.


In general, we can calculate the evolution using the dissipator in Eq.~(\ref{xii}). We can obtain the evolution using the dissipator given in Eq.~(\ref{xi}) just setting $\gamma_{3}=0$. The oscillation Hamiltonian in vacuum and in matter is taken in its diagonal form. Usually in vacuum $H_{S}$ is written in the mass basis as $H_{S}=diag\{E_{1},E_{2}\}$ and when the oscillation occurs in constant matter, it is possible to write the Hamiltonian as $H_{S}=diag\{\tilde{E}_{1},\tilde{E}_{2}\}$ using the effective mass basis. Note that we have defined two different subsystems of interest $S$, one for neutrinos in vacuum and another for neutrinos in constant matter, but both $H_{S}$ are diagonal.

We are going to use the approximation $E_{i}=E + m_{i}/2E$ and $\tilde{E}_{i}=E +\tilde{m}_{i}/2E$. The Eq.~(\ref{viii}) can be written as
\begin{equation}
\left(\begin{array}{ c } 
\dot{\rho}_{1}(x) \\
\dot{\rho}_{2}(x) \\
\dot{\rho}_{3}(x) \end{array} \right) = 
\left(\begin{array}{ c c c} 
-\gamma_{1} & -\Delta & 0\\
\Delta & -\gamma_{1} & 0\\
0 & 0 & -\gamma_{3} \end{array} \right)\left(\begin{array}{ c } 
{\rho}_{1}(x) \\
{\rho}_{2}(x) \\
{\rho}_{3}(x) \end{array} \right)\,,
\label{xiii}
\end{equation}
where $\Delta=\Delta m^{2}/2 E$. If the propagation is in matter, we can evoke the effective quantities, which are $\Delta\rightarrow \tilde{\Delta}=\Delta \tilde{m}^{2}/2E$, $\gamma_{i} \rightarrow \tilde{\gamma}_{i}$ by following the Eq. (\ref{viiextra}) and $\rho_{i} \rightarrow \tilde{\rho}_{i}$. Of course, this changes nothing from the point of view of the equation solution and from now on, we do not mention more this similarity. Further, the component $\rho_{0}$ has a trivial differential equation given by $\dot{\rho}_{0}(x)=0$  and its solution is $\rho_{0}(x)= \rho_{0}(0)$ that in two neutrino oscillation means $\rho_{0}(x)=1/2$.
%
%
The Eq. (\ref{xiii}) can be written in short form as
\begin{equation}
\dot{R}(t)= \mathbbm H R(t) \,,
\label{xiv}
\end{equation}
where the eigenvalues of $\mathbbm H$ are $\lambda_{0}=-\gamma_{3}$,  $\lambda_{1}=-\gamma_{1} - i \Delta$ and $\lambda_{2}=-\gamma_{1} + i \Delta $. For each eigenvalue it is possible to obtain a correspondent eigenvector, $\textbf{u}_{0}$, $\textbf{u}_{1}$, $\textbf{u}_{2} $ that compose the matrix $\mathbbm A =[\textbf{u}_{0}, \textbf{u}_{1}, \textbf{u}_{2} ] $ that diagonalizes the matrix $\mathbbm H$ by performing the following similarity transformation:  $\mathbbm A^{\dag} \mathbbm H \mathbbm A$. The solution of the Eq.~(\ref{xiv}) is given by
\begin{equation}
R(x)= \mathbbm M(x) R(0) \,,
\label{xv}
\end{equation}
where $\mathbbm M(x)$ is obtained making 
\begin{equation}
\mathbbm M(x)= \mathbbm A . diag\{e^{\lambda_{0}x},e^{\lambda_{1}x},e^{\lambda_{2}x} \} . \mathbbm A^{\dag} \,.
\label{xvi}
\end{equation}

Furthermore, it is useful to write the propagated state  which in this case is given by
\begin{equation}
\rho(x) = \left(\begin{array}{ c c } 
\rho_{0}(x)+\rho_{3}(x) & \rho_{1}(x)- i \rho_{2}(x)\\
\rho_{1}(x)+ i \rho_{2}(x) & \rho_{0}(x)-\rho_{3}(x)\,  \end{array} \right)\,.
\label{xvii}
\end{equation}

From the Eq. (\ref{xiii}), one can see that the propagated state is written as
\begin{equation}
\rho(x)=\left(\begin{array}{c c} 
\frac{1}{2}+\frac{1}{2}e^{-\gamma_{3} x}\cos 2\theta & \frac{1}{2}e^{-(\gamma_{1}-i\Delta)x}\sin 2\theta \\
\frac{1}{2}e^{-(\gamma_{1}+i\Delta)x}\sin 2\theta & \frac{1}{2}-\frac{1}{2}e^{-\gamma_{3} x}\cos 2\theta \\\end{array} \right)\,,
\label{xviii}
\end{equation}
where it is possible to identify two unusual behaviors. The off-diagonal entries are called coherence elements and it has a damping term that eliminates the quantum coherence during the propagation. This is the exact definition for decoherence effect and we can see clearly that such effect is associated with the matrix elements $\gamma_1$. The diagonal elements in Eq.~(\ref{xviii}) are known as population elements and they are related to the quantum probabilities of obtaining the eigenvalue $E_{1}$ or $E_{2}$ of the observable $H_{S}$. 

In the absence of dissipative effects, the observable is diagonal in the mass basis and the diagonal elements of the state are independent of the distance, but in the state in Eq.~(\ref{xviii}) the probability elements change with the propagation. This dissipative effect implies that the neutrinos may change their flavor without using the oscillation mechanism. As the asymptotic state is a complete mixing, the $\gamma_{3}$ in diagonal elements is called relaxation effect.  


The flavor oscillation probabilities can be obtained from the Eq.~(\ref{vii}) and $\rho^{f}_{11}$ element is the survival probability that is written as  
\begin{equation}
P_{\nu_{\alpha}\rightarrow\nu_{\alpha}} = \frac{1}{2}\bigg[1+e^{-\gamma_{3} x}\cos^{2}2\theta+ e^{-\gamma_{1} x}\sin^{2}2\theta\cos\left(\Delta x\right)\bigg]\,.
\label{xix}
\end{equation}

In Eq. (\ref{xix}), the asymptotic probability, $x\rightarrow \infty$, goes to a maximal statistical mixing, $ P_{\nu_{\alpha}\rightarrow\nu_{\alpha}}=1/2$, and this happens for any mixing angle. Thus, by means of this approach, the neutrino may change its flavor and it does not need to use the oscillatory mechanism to this end~\cite{ben, workneut}. In fact, while the decoherence effect, through $\gamma_{1}$ parameter, eliminates the oscillation term , the relaxation effect, $\gamma_{3}$ parameter, eliminates the term in the probability that depends only on the mixings.   

%
%

When the propagation is performed with the dissipator given in Eq.~(\ref{xi}), we obtain some important differences. In this case, $\mathbbm{H}$ has only two non-trivial eigenvalues  which are equal to $\lambda_{1}$ and $\lambda_{2}$ which were derivated before. Then, the matrix $\mathbbm{M}(x)$ is changed to
\begin{equation}
\mathbbm{M}(x) = \mathbbm{A} .diag \{1,e^{\lambda_{1} x},e^{\lambda_{2} x}\}. \mathbbm{A}^{\dag}\,,
\label{xx}
\end{equation}
and consequently, the state is written as 
\begin{equation}
\rho(x)=\left(\begin{array}{c c} 
\frac{1}{2}+\frac{1}{2}\cos^{2}\theta & \frac{1}{2}e^{-(\gamma_{1}-i\Delta)x}\sin 2\theta \\
\frac{1}{2}e^{-(\gamma_{1}+i\Delta)x}\sin 2\theta & \frac{1}{2}-\frac{1}{2}\cos^{2}\theta \\\end{array} \right)\,.
\label{xxi}
\end{equation}

In the state above, there is only influence of the decoherence effect and only the coherent elements are eliminated during the propagation. In this case, the survival oscillation probability is written as
\begin{equation}
P_{\nu_{\alpha}\rightarrow\nu_{\alpha}}	 = 1 - \frac{1}{2}\sin^{2}(2\theta)\Big[1-e^{-\gamma_{1} x}\cos(\Delta x)\Big]\,.
\label{xxii}
\end{equation}

This probability was discussed in Refs.~\cite{ ben,workneut,  lis} only in the vacuum approach, but we are showing that when the open quantum system approach is applied carefully a similar probability is obtained for the propagation in matter as well.

Then, when there is energy conservation in subsystem of interest, $[H_{S},V_{k}]=0$, the asymptotic probability, $x\rightarrow \infty$, still depends on the mixing angle as 
\begin{equation}
P_{\nu_{\alpha}\rightarrow\nu_{\alpha}}	 = 1 - \frac{1}{2}\sin^{2}(2\theta)\,.
\label{xxiiextra}
\end{equation}
%


In this approach, the dynamics is made through Eq.~(\ref{ii}) and it depends on how the subsystem of interest interacts with the environment following constraint: $[H_{S},V_{k}]=0$ or $[H_{S},V_{k}]\neq0$. 

From a mathematical point of view, when we consider neutrinos like an open quantum system and taking into account the considerations explored in this section, one can see that there are not significant differences in deriving the quantum evolution in vacuum or in constant matter. This result is trivial in closed approach, but it is not a trivial result in this open approach. In fact, the similarity between these two propagation conditions in open approach is only true when the reservoir interacts in some way with the subsystem of interest represented by $S$ that here, it was defined using mass state in vacuum propagation or effective mass state in matter propagation. Otherwise, there will not be similarities between the vacuum and matter propagation \cite{ ben1, fo}.

\section{Neutrinos in Non-Uniform Matter}

In many situations the neutrino propagation occurs where the matter density is not constant. We are going to assume  neutrino evolution in  non-constant matter only in situations where the adiabatic limit is valid~\cite{kim,moh}. Thus, the results obtained in this situation are similar to those obtained for propagation in constant matter. The main focus now is to understand which dissipative effects act on neutrinos supposing that the source is far away from the Earth. Solar neutrinos are a great example that we want to study. 

Using the same point of view from the previous section, we can write a diagonal Hamiltonian using the effective mass basis. We start with the quantum dissipator written in Eq.~(\ref{xii}). Thus, we have to solve the same evolution equation given in Eq.~(\ref{xiii}), but on the right side, the elements of the first matrix are distance dependent as well. So, the Eq.~(\ref{xiv}) is written now as
\begin{equation}
\dot{R}(x)= \mathbbm H(x) R(x) \,,
\label{xxiii}
\end{equation}
and it  has a solution similar to Eq. (\ref{xv}), but $\mathbbm{M}(x)$ is proportional to
\begin{equation}
\mathbbm M(x)\propto diag\{e^{\int^{R_{\odot}}_{r}\lambda_{0}(x)dx},e^{\int^{R_{\odot}}_{r}\lambda_{1}(x)dx},e^{\int^{R_{\odot}}_{r}\lambda_{2}(x)dx} \} \,,
\label{xxiv}
\end{equation}
where $r$ and $R_{\odot}$ are the creation and detection point, respectively. As $\mathbbm{A}$ is defined in the same way of the previous section, the $\lambda_{i}(x)$ has the same form of $\lambda_{i}$ defined in Eq. (\ref{xiv}), but here $\Delta \rightarrow \tilde \Delta (x)$ and $\gamma_{i} \rightarrow \tilde{\gamma}_{i}$ may depend on distance. Even for $ \lambda_{0}$ the distance dependence may exist~\cite{fo}.

Notice that the energy conservation is given by $[H_{S}, V_{k}]=0$, but in general the $H_{S}$ in vacuum propagation is different from $H_{S}$ in matter propagation. Consequently, when one imposes energy conservation in matter propagation there is not energy conservation in vacuum propagation and vice-versa. On the other hand, it is possible to obtain a model where the energy conservation is always kept even when $H_{S}$ in vacuum and in matter propagation are different. In this case, the dissipative quantum operator has a distance dependence such that $V_{k}$ changes to $V_{k}(x)$ and can be written as
\begin{equation}
V_{k}=V_{k}(x)=\left(\begin{array}{c c} 
2\sqrt{\gamma_{1}}\cos[\Theta(x)] & \sqrt{\gamma_{1}}\sin[\Theta(x)] \\
\sqrt{\gamma_{1}}\sin[\Theta(x)]& 0 \\\end{array} \right)\,,
\label{pi}
\end{equation}
where $\Theta(x)=2(\theta-\tilde{\theta}(x))$ and the effective angle is given by
\begin{align}
\tilde{\theta}(x)&=\frac{1}{2}\arcsin \left(\sqrt{\frac{\Delta^{2} \sin^{2}[2\theta]}{(\Delta \cos[2\theta]-A(x))^{2}+\Delta^{2} \sin^{2}[2\theta]}}\right) \,.
\label{pii}
\end{align}

The off-diagonal elements in vacuum case are null and the element $\{V_{k}(x)\}_{11}=2\sqrt{\gamma_{1}}$ such that the quantum dissipator in Eq. (\ref{xi}) is not changed. Supposing the adiabatic limit or constant density matter, we can rewrite the evolution in mass basis into  effective mass basis where one considers the addition of  the potential matter. In this case, the dissipation operator $V_{k}(x)$ in vacuum changes to $\tilde{V}_{k}(x)=\tilde{U}^{\dag} U V_{k} U^{\dag}\tilde{U}$ in matter propagation, such that it is written as
\begin{equation}
\tilde{V}_{k}(x)=\left(\begin{array}{c c} 
2\sqrt{\gamma_{1}}\cos^{2}[\Theta(x)] &0 \\
0& -2\sqrt{\gamma_{1}}\sin^{2}[\Theta(x)]\\\end{array} \right)\,,
\label{piii}
\end{equation}
and the dissipator in Eq. (\ref{xi}) continues unchanged as well. 

Thus, disregarding models where the operator in Eq. (\ref{pi}) differs by a unitary matrix, this is a unique model where energy conservation constraint in matter propagation and in vacuum propagation are satisfied simultaneously. This occurs due to the fact that energy conservation in matter propagation is given by $[\tilde{H}_{S}(x), \tilde{V}_{k}(x)]=0$, and this result is valid for any choice of matter potential.

So, as we can mentioned before, if we want that the evolution is purely decoherent, i. e., that the energy conservation, $[\tilde{H}_{S}(x), \tilde{V}_{k}(x)]=0$, is satisfied during the propagation even when the density matter varies, we must have a dissipation operator like the one in Eq. (\ref{pi}), because it takes into account how much the matter effect could change it.

%
%

Returning to the evolution given by Eq. (\ref{xxiii}), the state evolved using Eq. (\ref{xxiii}) is written as   
\begin{equation}
\tilde{\rho}_{m}(x)=\left(\begin{array}{c c} 
\frac{1}{2}+\frac{1}{2}e^{-\Gamma}\cos 2\tilde{\theta} & \frac{1}{2}e^{-\Gamma_{1}}\sin 2\tilde{\theta} \\
\frac{1}{2}e^{-\Gamma^{*}_{1}}\sin 2\tilde{\theta} & \frac{1}{2}-\frac{1}{2}e^{-\Gamma}\cos 2\tilde{\theta} \\\end{array} \right)\,,
\label{xxv}
\end{equation}
where we have defined
\begin{equation}
\Gamma=-\int^{R_{\odot}}_{r}\tilde{\gamma}_{3}(x)dx\,
\label{xxvi}
\end{equation}
and
\begin{equation}
\Gamma_{1}=-\int^{R_{\odot}}_{r}\tilde{\gamma}_{1}(x)dx +i \int^{R_{\odot}}_{r}\tilde{\Delta}(x)dx  \,,
\label{xxvii}
\end{equation}
where $\tilde{\gamma}_{1}(x)=\gamma_{1}$ if we consider the dissipation operator in Eq. (\ref{pi}).
    
In general, the second term in Eq.~(\ref{xxvii}) gives rise to fast oscillation terms in the off-diagonal elements and it is usually averaged out. Thus, the state has the following form
\begin{equation}
\tilde{\rho}_{m}(x)=\left(\begin{array}{c c} 
\frac{1}{2}+\frac{1}{2}e^{-\Gamma}\cos 2\tilde{\theta} & 0 \\
0 & \frac{1}{2}-\frac{1}{2}e^{-\Gamma}\cos 2\tilde{\theta} \\\end{array} \right)\,,
\label{xxviii}
\end{equation}
where, we conclude that in general we cannot have information about the decoherence effect  in this situation.

To obtain the usual adiabatic probability we use the fact that the effective mixing angle changes during the neutrino propagation and then, the mixing angle in the detection point must be different. We define the initial mass state from the Eq.~(\ref{vii}), where in the creation point, we used the effective mixing angles written as $\tilde{\theta}$. Then, we can change the representation by applying another mixing matrix with another mixing angle. Defining these angles in detection point as $\tilde{\theta}_{d}$, we have
\begin{equation}
\rho_{f}(x)=U_{d}\tilde{\rho}_{m}(x)U^{\dag}_{d}\,,
\label{xxix}
\end{equation}
where $U_{d}$ is the usual mixing matrix, but with mixing angle $\tilde{\theta}_{d}$. Then, the adiabatic survival probability, $\rho^{f}_{11}(x)$, is given by
\begin{equation}
 P^{adiab.}_{\nu_{e}\rightarrow\nu{e}}=\frac{1}{2}+\frac{1}{2}e^{-\Gamma}\cos2\tilde{\theta}\cos2\tilde{\theta}_{d}\,.
\label{xxx}
\end{equation}

In the survival probability above, if $\Gamma=0$, we recover the usual survival probability in the adiabatic limit case \cite{kim, moh}. The dissipation operator in Eq. (\ref{pi}) is obtained when the energy constraint, $[H_{S},V_{k}]=0$ is imposed and hence only decoherence effect might be described by $\tilde{\gamma}_{1}$ using the operator in Eq.~(\ref{xi}). However, the state more general for solar neutrinos does not hold the $\tilde{\gamma}_{1}$ in its description and then, we can conclude that quantum decoherence cannot be limited by solar neutrinos in general. On the other hand, as only $\tilde{\gamma}_{3}$  remains in the state (\ref{xxviii}) and in the probability (\ref{xxx}), in general, just the relaxation effect can be limited when one considers solar neutrinos

Now we analyze a situation mentioned in the subsection \ref{s} that is, for example, the same supposition that the authors in Ref. \cite{fo} used to put limit on decoherence effect using solar neutrinos.

So, we assume neutrinos propagate in matter in the situation where the adiabatic limit is satisfied. As usual, the Hamiltonian is $H_{S}=H_{osc}+H_{mat}$, where $H_{osc}$ is the oscillation Hamiltonian in vacuum and $H_{mat}$ is the matter potential. In addition, we assume energy conservation with two different conditions. One of them is when we suppose energy conservation only with the vacuum piece, $[H_{osc},\bar{V}_{k}]=0$, and  another one is when we assume energy conservation only with the matter potential piece, $[H_{mat},V'_{k}]= 0$. Note that the $\bar{V}_{k}$ and $ V'_{k}$ follow the definition given by in Eq. (\ref{viiextra}) and both of them are different of $ V_{k}$ that may commutate with $H_{S}$.  

These two situations can try to investigate only the decoherence effect. One of them the neutrino state in vacuum can be changed due to the decoherence effect even it is present in the Sun, for instance. With another one, it is possible to study decoherence effect in the Sun environment in order to change the matter effect through a dissipative phenomenon. 

As the energy conservation constraint in subsystem of interest was assumed whatever the place that neutrino will go through, for both situations the quantum dissipator used in the propagation in Eq. (\ref{v}) is given by Eq. (\ref{xi}). However, as we have mentioned, this quantum dissipation includes only quantum decoherence effect in the propagation. So, for the quantum evolution in both situations, the Eq.~(\ref{xxiii}) with $\mathbbm{H}$ is now given by

\begin{equation}
\mathbbm{H}=\left(
\begin{array}{ccc}
 -\gamma_{1}  & -\Delta -A \cos 2\theta & 0 \\
  \Delta +A \cos2 \theta & -\gamma_{1}  & -A \sin2 \theta \\
 0 & A \sin 2 \theta  & 0
\end{array}
\right)\,,
\label{xxxi}
\end{equation}
where $\gamma_{1}$ comes from $D_{mn}$ in Eq. (\ref{ix}) for both cases and in the equation above, $\mathbbm{H}$ was written in mass basis representation.

The characteristic polynomial of the above matrix has a complicated solution, but if we consider $\gamma_{1}$ is small such that it can be treated like a perturbation, we obtain in first order approximation the following eigenvalues:  
\begin{eqnarray}
\lambda_{0}=-\gamma_{1}\frac{A^{2}}{\Delta^{2}}\sin^{2} 2\tilde{\theta};\nonumber\\
\lambda_{1}=-\gamma_{1}+\gamma_{1}\frac{A^{2}}{\Delta^{2}}\sin^{2}2\tilde{\theta}-i\tilde{\Delta};\nonumber\\
\lambda_{2}=-\gamma_{1}+\gamma_{1}\frac{A^{2}}{\Delta^{2}}\sin^{2}2\tilde{\theta}+i\tilde{\Delta}.
\label{xxxii}   
\end{eqnarray}
where $A = \sqrt{2} G_{F} n_{e}$ and, for sake of simplicity, we can rewrite $\mathbbm{H}$ in the effective mass basis, such that we get
\begin{equation}
\mathbbm{H}=\left(
\begin{array}{ccc}
 -\tilde{\gamma}_{1}  & -\tilde{\Delta}  & 0 \\
  \tilde{\Delta}  &  -\tilde{\gamma}_{1} & 0 \\
 0 & 0  & -\tilde{\gamma}_{3}
\end{array}
\right)\,,
\label{xxxiii}
\end{equation}
with $\tilde{\gamma}_{3}=\gamma_{1} A^{2}\sin^{2}2\tilde{\theta}/\Delta^{2}$ and $\tilde{\gamma}_{1}=\gamma_{1}-\tilde{\gamma}_{3}$. From $\mathbbm{H}$ given by Eq.~(\ref{xxxiii}) we obtain the same state that was given in Eq.~(\ref{xxv}) where $\Gamma_{1}$ would be defined by $\tilde{\gamma}_{1}$ while $\Gamma$ by $\tilde{\gamma}_{3}$. With the same arguments that was given before, $\Gamma_{1}$ becomes null and we obtain the state in Eq. (\ref{xxviii}).  The interpretation is similar that was done before where $\Gamma_{1}$ is not important and only the relaxation effect, $\Gamma \propto \tilde{\gamma}_{3}$, may change the probability.

In these two situations the constraints are $[H_{S},\bar{V}_{k}]\neq 0$ and $[H_{S},V'_{k}]\neq 0$. Thus, we could expect that the result for these different constraints,  $[H_{osc},\bar{V}_{k}]=0$ and $[H_{mat},V'_{k}]= 0$, are obtained by  an evolution using the dissipator in Eq. (\ref{xii}), as we have seen in subsection \ref{s}. Besides, this result show that there is not a way to separate the subsystem of interest $S$ in pieces which may or may not interact with the environment and here, as we have $[H_{S},\bar{V}_{k}]\neq 0$ and $[H_{S},V'_{k}]\neq 0$ the relaxation effect appears naturally.

The decoherence and relaxation effects when the propagation in matter may have different magnitude from the vacuum case. However, independently of we assume $[H_{osc},\bar{V}_{k}]=0$ or  $[H_{mat},V'_{k}]= 0$, we have the same result for the dissipative effects. This looks like an apparent problem because we cannot differentiate between these dissipative models in the solar neutrino case, for example.  

In special, the case where $[H_{osc},\bar{V}_{k}]=0$  the Eq. (\ref{xxxiii}) shows relaxation effect is proportional to the decoherence effect for neutrinos propagating in vacuum (the same occurs for the case $[H_{mat},V'_{k}]= 0$ \cite {bur}). This was the result obtained by Ref. \cite{fo} and thus, from this model-dependent approach, the decoherence effect in vacuum, $\gamma_{1}$, was limited by authors in Ref. \cite{fo}. Besides, the $\bar{V}_{k}$ wrote there in our notation is written as
\begin{equation}
\bar{V}_{k}=\left(\begin{array}{c c} 
2\sqrt{\gamma_{1}} &0 \\
0& 0\\\end{array} \right)\,.
\label{piv}
\end{equation}
which is different from the $V_{k}(x)$ given in Eq. (\ref{pi}), where the matter potential becomes important and the energy conservation is always satisfied even when the propagation is through in non-constant matter. 	 

In the Ref. \cite{bur} the authors made a microscopic model to the interaction between neutrinos and the solar environment and they reached a dynamic equation similar to Eq. (\ref{xxxii}), but there the dissipation effect appears as a consequence of this microscopic model where $[H_{mat},V'_{k}]= 0$ was satisfied. The dynamic obtained in Ref. \cite{fo} was also obtained by authors in Ref. \cite{bur} even the study propose being different one another, of course, they reached to same probability as well.  

Therefore, the result of the last example is interesting because it has not trivial interpretation. And there is not in the literature a reliable limit for decoherence effect in the channel $\nu_{e}\rightarrow \nu{_\mu}$ obtained from a model-independent approach. Surely, it exists only limits on the relaxation and decoherence effects in the case of a particular model-dependent approach used by Ref. \cite{fo} in two neutrino approximation. So, other analysis using a general model-independent approach can be done using neutrinos that come from other sources, where the constraint $[H_{S},V_{k}]=0$ can without any doubt be satisfied and the decoherence effect be limited.

\section{Comments and Conclusion}

The quantum dissipator in Eq.~(\ref{xi}) is related to decoherence effects while the quantum dissipator in Eq.~(\ref{xii}) is related to decoherence plus relaxation effects. We explicitly relate decoherence effects with a quantum dissipator that conserves energy in the subsystem of interest, a condition that is fulfilled if $[H_{S},V_{k}]=0$. If such condition is violated, then we relate such quantum dissipator with relaxation effects. So, we introduce the unique form in which this condition is satisfied in all points of the evolution since $H_{S}$ is the Hamiltonian that governs the evolution in the usual approach.  This means that $H_{S}$ is composed by mass and interaction Hamiltonians in matter propagation and only mass Hamiltonian in the case of the vacuum propagation.

We emphasized the differences and similarities between the $\mathbbm{H}$ eigenvalues that are obtained when we used the dissipators in Eqs.~(\ref{xi}) and~(\ref{xii}). We clearly see when the relaxation effect is present in the model and how the behavior of the states is changed in the situation with and without the relaxation effects. We discussed the neutrino evolution in vacuum and in matter with constant density and pointed out how these situations can have similar treatments in open quantum system formalism. We showed that in general the probabilities in vacuum and in constant matter can be written in similar ways, which is not an obvious result in this approach. It is interesting to note that through the model developed in this article, we do not need to use any method of approximation to obtain the probabilities in all cases. This is different from what we can find in the literature \cite{ben1,fo, bur}.  

We analyzed also the situation where the matter density is not constant. We obtained a dissipation operator in Eq. (\ref{pi}) that conserves  energy  during the neutrino propagation through a variable matter density.  We showed that the decoherence effect from our model-independent analysis cannot be limited in situations where experiments can no longer access the oscillation term in the  probabilities, as it is the case when the source is very far away from the detection point. On the other hand, the relaxation effect may still be tested and limited in such situations. Although, as it was made in Ref \cite{fo} through a model-dependent approach, it is possible to limit the decoherence in this case because the decoherence effect is connected in some way with the relaxation effect.  However, as we have pointed out, the relaxation and decoherence effects are different phenomena and both bring different behavior to the neutrinos. 

We identified some ambiguities in the definition of decoherence effects present in the literature~\cite{fo}, where there is no clear distinction between decoherence and relaxation effects. In our understanding, the term {\it decoherence} is often used to describe a combined effect of decoherence and relaxation when neutrino evolves in a medium with variable density.  We described how it would be a dissipative model with only quantum decoherence effects for propagation in matter with non-constant density. From the dissipative operator obtained in Eq. (\ref{pi}), it was possible to see why the decoherence effect was limited in Refs. \cite{fo} and mentioned in Ref. \cite{bur}. In fact, in those cases it could not exist decoherence effect only, but another effect related in some way with the decoherence effect, because the dissipative operator used by these references, Eq. (\ref{piv}), violates the condition $[H_{S},V_{k}]=0$, when neutrinos propagate in constant or non-constant matter.

Comparing our approach with the ones found in the literature, it is possible to conclude that it avoids all the ambiguities about which kind of dissipative effect is acting on neutrinos. As stated before,  the limit for the decoherence effect should be obtained through experiments that access the oscillation pattern in the flavor neutrino probabilities, like KamLAND~\cite{balieiro}, for instance. The result of the Ref.~\cite{fo} can be interpreted as an upper  limit on the decoherence effect which comes, in fact, from the restriction on the relaxation effect, once that both effects are connected in this model dependent analysis. Our model independent approach is able to put bounds on all dissipation effects in a direct way.

\section{Acknowledgments}
We would like to thank CNPq and FAPESP for several financial supports.
R.L.N.O. is grateful to L. Ostrar. and J.A.B. Coelho for instructive discussions and thanks for the support of funding grants 2012/00857-6 and 2013/11651-2 São Paulo Research Foundation (FAPESP).



%

\begin{thebibliography}{100}
\bibitem{escrever}
We are making analyses to put bounds on relaxation for solar neutrinos. This analysis will appear in elsewhere, but our preliminar results are $\gamma<{10^{-18}}$ eV at $95\%$ at C.L. for relaxation effect. The result constrains enormously the relaxation effect, such that it becomes unreachable to sensibility of any terrestrial experiments.
\end{thebibliography}






\end{document}